\documentclass[PRB,twocolumn,showpacs,amsmath,superscriptaddress]{revtex4-1}
\usepackage[colorlinks,urlcolor=blue,linkcolor=blue,citecolor=blue]{hyperref}
\usepackage{overpic}
\usepackage{subfigure}
\usepackage{latexsym}
\usepackage{braket}
\usepackage{amsfonts}
\usepackage{color,xcolor}
\usepackage{bm}
\usepackage{array}
\usepackage{booktabs}
\usepackage{caption}
\usepackage{multirow}

\makeatletter

\newcommand{\Rmnum}[1]{\expandafter\@slowromancap\romannumeral #1@}

\begin{document}

	\title{Magnetic flux induced topological superconductivity in magnetic atomic rings}

	\author{Jinpeng Xiao}
	\affiliation{
		School of Mathematics and Physics, Jinggangshan University, Ji'an 343009, P. R. China}
	\author{Qianglin Hu}
    \affiliation{
		Department of Physics and Electronics Engineering, Tongren University, Tong Ren 554300, P. R. China}
	\author{Xiaobing Luo}
   \altaffiliation{Corresponding author: xiaobingluo2013@aliyun.com}

   \affiliation{Department of Physics, Zhejiang Sci-Tech University, Hangzhou 310018, P. R. China}
	
	\date{\today}
	
	\begin{abstract}
There have been numerous studies on topological superconductivity in magnetic atomic chains deposited on s-wave superconductors. Most of these investigations have focused on spin-orbit interactions or helical spin orders. In this paper, we propose a model for achieving one-dimensional topological superconductivity in a magnetic atomic ring. This model utilizes a magnetic field and an antiferromagnetic/ferromagnetic order, under the condition that the magnetic field is perpendicular to the moments of the magnetic order. On a quasi-one-dimensional substrate surface, where the half-filled ring favors an antiferromagnetic configuration, we demonstrate that either the magnetic field itself or a Rashba spin-orbit coupling guarantees the perpendicularity. On a two-dimensional surface, where the ring favors ferromagnetic orders, the perpendicularity is achieved by introducing a minor Rashba spin-orbit coupling.

	\end{abstract}
	\maketitle
	\section{Introduction}\label{Sec1}
The exploration of topological superconductors, which are theoretically predicted to host Majorana zero modes\cite{24,25,26,27,28,29,30,31,32,33,34,35,36,37,38}, has emerged as a burgeoning field of research over the past decade, wherein magnetism plays crucial roles. Typically, an s-wave superconductor exhibits antagonistic relationships between magnetism and superconductivity. However, carefully engineered artificial structures enable the coexistence of magnetism and superconductivity at the magnet-superconductor interfaces through the proximity effect\cite{75}.
Recent years have witnessed extensive research on realizing topological superconductivity by employing the proximity effect. Numerous studies have focused on one-dimensional magnetic atomic chains accompanied by helical\cite{16,21,22,23,39,40,41,42,43,44}, antiferromagnetic (AFM)\cite{46,76,48,49,50,51,52}, and ferromagnetic (FM) orders\cite{45,47}. Notably, compared to other magnetic orders, an AFM order does not lift Kramer's degeneracy between opposite spins, rendering it more compatible with spin-singlet superconductivity\cite{6,9,10,11,12}, and thus more suitable for engineering topological superconducting states.

For a long time, considerable research endeavors have been devoted to AFM materials, which possess interesting features such as robustness against magnetic perturbations, absence of stray fields and exhibition of ultrafast dynamics\cite{1,2,3,4,5,6,7,8}. Numerous methods have been explored and developed to achieve AFM ordering in materials. In low dimensional systems, a prevalent method is to utilize the proximity effect, wherein the AFM order is induced in magnetic atoms when they are deposited on the surface of strong AFM materials such as $Mn_{2}C$, $NiPS_{3}$, $FeB$, or $MnB$\cite{18,19,20}. Scanning tunneling microscope (STM) offers an alternative option to obtain AFM orders as it can manipulate atoms one by one. Various magnetic orders have been achieved on superconductor substrates through artificial construction of magnetic atoms via STM\cite{13,14,15,16,45}.
In the last decade, an intriguing method has been proposed for quasi-one-dimensional systems, where a magnetic atomic chain can self-organize into a helical order with the pitch angle equal to $2k_{F}a$ by employing the Ruderman-Kittel-Kasuya-Yosida (RKKY) mechanism\cite{21,22,23}. As a consequence, the magnetic chain exhibits AFM ordering when the system is half-filled.

Recently, a method has been developed to explore topological superconducting states in quasi-one-dimensional systems by exploiting a magnetic flux. When the flux threads through the loop of an Aharonov-Bohm interferometer\cite{53,54,55,56,57,58} or the core of a nanotube\cite{59,60,61,77}, a topological superconducting state can be induced in the presence of a spin-orbit coupling (SOC). Nevertheless, the majority of existing routes toward realizing topological superconductivity in quasi-one-dimensional systems rely on two basic ingredients: strong spin-orbit couplings or helical magnetic orders.
In this paper, we propose a topological superconducting model consisting of a one-dimensional AFM atomic ring deposited on the surface of an s-wave superconductor, in the absence of SOCs. The ring exhibits superconductivity due to the proximity effect from the substrate. We find that topological superconductivity emerges when a magnetic flux threads through the ring and when the AFM moments are perpendicular to the external magnetic field. Further studies reveal that the perpendicularity is guaranteed by either the external magnetic field or a Rashba SOC. Therefore, the proposed model serves as a promising platform for realizing topological superconductivity.

The paper is organized as follows: In Sec. \ref{Sec2}, we introduce the model of the antiferromagnetic ring and discuss its topological properties. In Secs. \ref{Sec3} and \ref{Sec4}, we examine the influences of the Rashba spin-orbit coupling and magnetic field, respectively. Sec. \ref{Sec5} discusses the magnetic order when both the Rashba spin-orbit coupling and magnetic field coexist with superconducting pairing. In Sec. \ref{Sec6}, we extend the study to a ring deposited on a two-dimensional surface, where the system favors a ferromagnetic order. Staggered arrangements of magnetic atoms restore topological superconductivity, and the orientation of the magnetic moments is determined by introducing a minor Rashba spin-orbit coupling. Finally, in Sec. \ref{Sec7}, we provide a brief summary and discussion of the results.
	
	\section{Model of antiferromagnet ring adopt on s-wave superconductor}\label{Sec2}
\begin{figure}
\scalebox{1.0}{\includegraphics[width=0.45\textwidth]{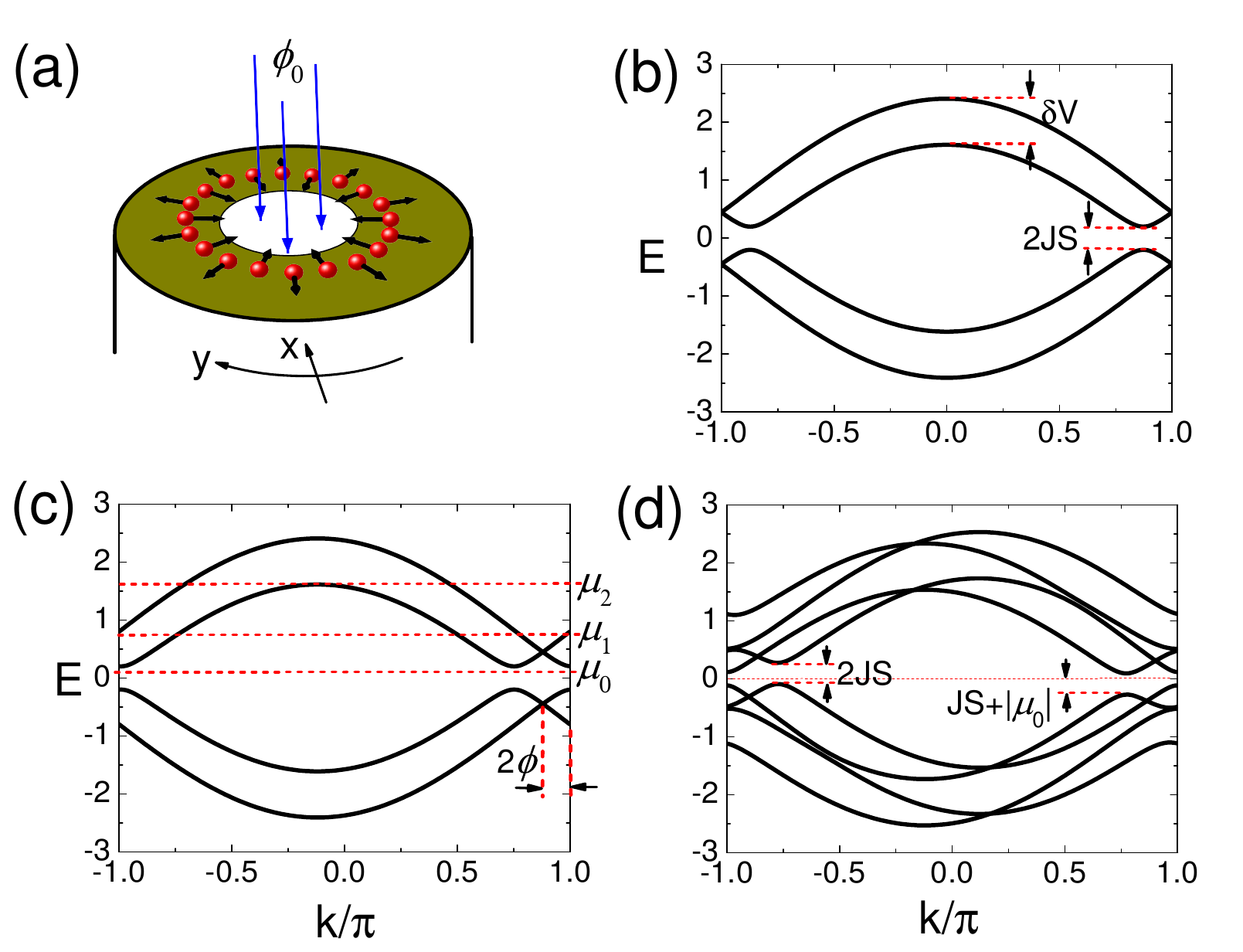}}
\caption{\label{fig1} (a) A schematic representation of an AFM ring with a flux threading through it. (b),(c) The normal-state bands of the AFM ring when the chemical potential $\mu=0$. In (b), the phase $\phi$ $(=\phi_{0}/L$, where $L$ is the number of sites in the ring) is 0, while in (c), $\phi$ is $0.06\pi$. The magnetic Zeeman field $V$ splits the bands of different spins by an energy $\delta V$. The phase $\phi$ induces a shift of the bands along the wave vector $k$. (d) The bands of Hamitonian (\ref{eq2}) with $\mu=\mu_0=0.1$, $\Delta=0.3$ and $\phi=0.06\pi$. The other parameters in (b)-(d) are $JS=0.2$ and $V=0.4$.}
\end{figure}

We consider a ring of antiferromagnetically ordered magnetic atoms deposited on the top surface of a hollow cylinder s-wave superconductor, as illustrated in Fig. \ref{fig1}(a). The underlying superconductor induces superconductivity in the magnetic ring due to the proximity effect\cite{21}. When a perpendicular magnetic field is applied on the surface, it creates both a Zeeman field $V$ on the ring and a flux $\phi_{0}$ through the ring. Suppose the surface is in the $x$-$y$ plane, as the magnetic moments lie in the surface, the effective Hamiltonian of the ring can be expressed as
\begin{eqnarray}\label{eq1}
H_{0}&=&\sum_{j}te^{-i\phi}c_{j}^{\dag}c_{j+1}+J(-1)^{j}c_{j}^{\dag}\mathbf{S}_{j}\cdot\boldsymbol{\sigma}c_{j}\nonumber\\
&+&\mu c_{j}^{\dag}c_{j}+V c_{j}^{\dag}\sigma_z c_{j}+\Delta c_{j\uparrow}^{\dag}c_{j\downarrow}^{\dag}+h.c.
\end{eqnarray}
The first term is the hopping term with a phase factor $\phi=\phi_{0}/L$, which can also be effectively obtained through a supercurrent\cite{51,78,79}. $t$ is the itinerant electrons' hopping amplitude between nearest-neighbor sites and $L$ is the total number of sites in the ring. Here, $c_{j}^{\dag}=(c_{j\uparrow}^{\dag},c_{j\downarrow}^{\dag})$ with $c_{j\sigma}^{\dag}$ the electron creation operators on site $j$. The second term accounts for the in-plane AFM arranged magnetic atoms with $\mathbf{S}_{j}=S(\cos\varphi,\sin\varphi,0)$, where $\varphi$ is a random angle. $J$ stands for the exchange coupling constant between itinerant electrons and the onsite magnetic moment, and $\boldsymbol{\sigma}=(\sigma_x,\sigma_y,\sigma_z)$ is the vector of spin Pauli matrices. The third and fourth terms represent the chemical potential and external Zeeman field, respectively. The last term is the pairing term due to the proximity effect, and $\Delta$ is the induced pairing parameter which is assumed to be uniform. All the energies are in units of $t$ in the entire paper.

\begin{figure}
\scalebox{1.0}{\includegraphics[width=0.45\textwidth]{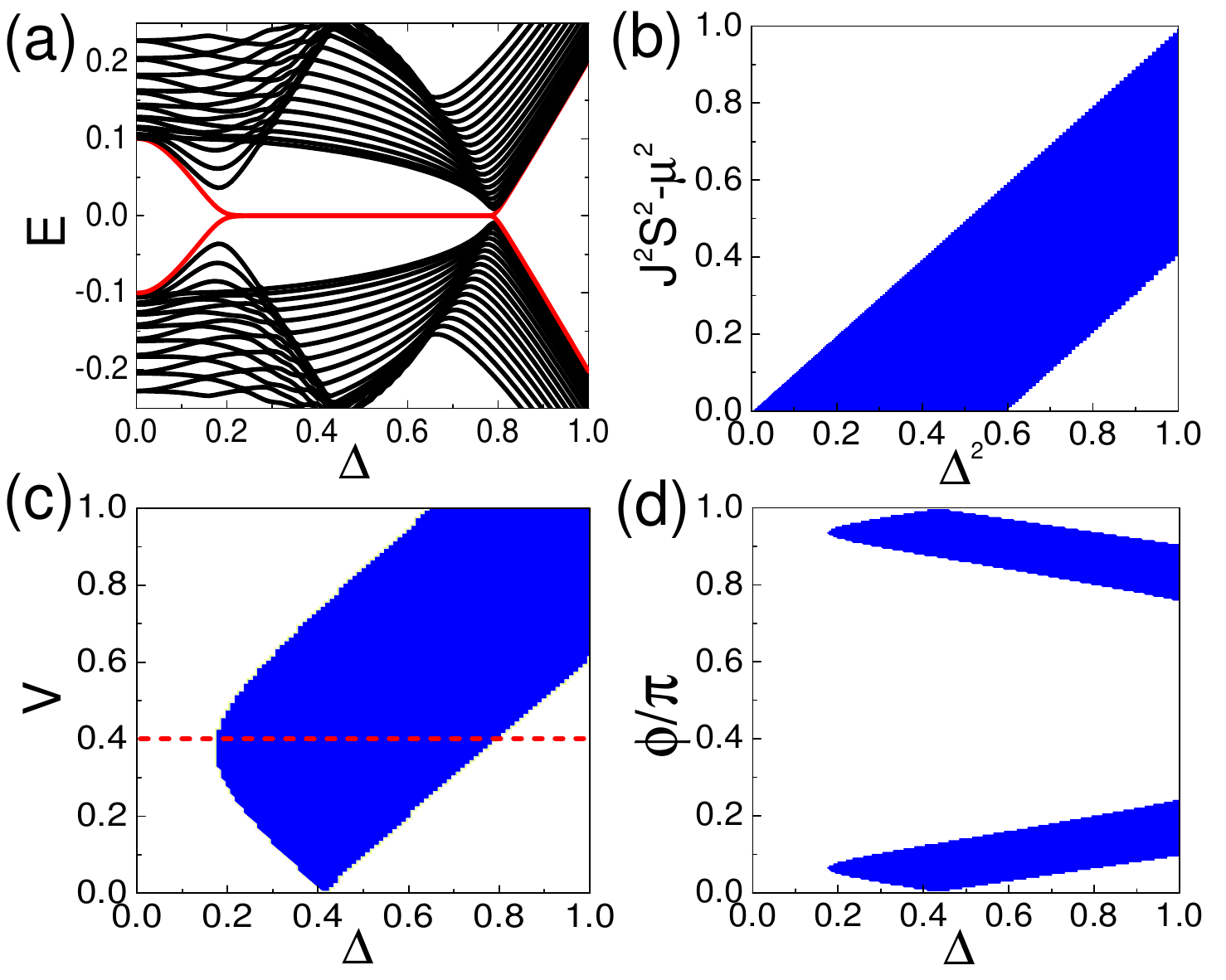}}
\caption{\label{fig2}  (a) The open boundary spectrum of the AFM ring with 200 sites as a function of the superconducting pairing strength $\Delta$. (b)-(d) The topological phase diagrams for the ring. In (a)-(c), the phase $\phi$ is set to $0.06\pi$. In (a),(b),(d) $V=0.4$; (a),(c),(d) $\mu=0.1$, $JS=0.2$. The blue regions represent topologically nontrivial phases. The red dashed line in (c) marks the scanning trajectory of the open boundary spectrum shown in (a).}
\end{figure}

The Hamiltonian can be expressed in momentum space as $H_{0}=\sum_{k}\Psi_{k}^{\dag}H_{0}(k)\Psi_{k}$, where the basis spinor is $\Psi_{k}=[f_{A},f_{B}]^{T}$ with $f_{\delta}=[c_{\delta k\uparrow},c_{ \delta k\downarrow},c_{\delta -k\downarrow}^{\dag},-c_{\delta -k\uparrow}^{\dag}]$. Here, $c_{ \delta k\alpha}=\sqrt{\frac{2}{L}}\sum_{j}e^{-ikR_{j\delta}}c_{ \delta j\alpha}$, where $\alpha$ labels spins and $\delta=A/B$ represents the sublattice sites.
\begin{eqnarray}\label{eq2}
H_{0}(k)&=&[\xi_{0}(k)\tau_z+\eta_{0}(k)]\Gamma_{k/2}+\Delta\tau_{x}+V\sigma_z\nonumber\\
&+&\mu\tau_z+(Js_{x}\sigma_x+Js_{y}\sigma_y)\Gamma_z,
\end{eqnarray}
where $\xi_{0}(k)=2t\cos\phi\cos\frac{k}{2}$, $\eta_{0}(k)=2t\sin\phi\sin\frac{k}{2}$, $s_{x}=S\cos\varphi$ and $s_{y}=S\sin\varphi$. $\tau$ are Pauli matrices acting on particle-hole space. $\Gamma_{k/2}=\cos\frac{k}{2}\Gamma_{x}-\sin\frac{k}{2}\Gamma_{y}$ with $\Gamma_{x,y,z}$ being three pauli matrices acting on the sublattice space.
The normal-state energy dispersions are given by $\varepsilon_{1,2}^{\pm}(k)=\mu\pm\sqrt{[\xi_{0}(k)+\eta_{0}(k)\mp V]^{2}+J^{2}S^{2}}$, which are illustrated in Figs. \ref{fig1}(b) and \ref{fig1}(c). The presence of the Zeeman field causes a splitting of the bands for different spins. The flux $\phi_{0}$ induces a shift of the dispersions by $2\phi$ along the wave vector $k$, leading to an odd number of Fermi-level crossings on each side of the first Brillouin zone when the chemical potential $\mu$ satisfies $JS<|\mu|<\mu_{1}$ or $\mu_{2}<|\mu|<\mu_{2}+\delta V$. According to Kitaev's criterion\cite{62}, the system becomes topologically nontrivial when a weak superconducting pairing is introduced in these regions. However, as illustrated in Fig. \ref{fig1}(d), the bands exhibit asymmetric behavior between the two sides of the first Brillouin zone. Consequently, the condition $|\mu|<JS$ becomes a prerequisite for the system to be gapped, contradicting the aforementioned Kitaev's criterion. This observation suggests that the topological criterion cannot be met in a gapped system. Interestingly, when considering the superconducting pairing, the bands exhibit a twisting behavior near the Fermi level, indicating that the system may become nontrivial if the pairing strength exceeds the energy gap. Figure \ref{fig2}(a) illustrates the open boundary energy spectrum with varying pairing strength $\Delta$. Zero energy modes (denoted by red solid lines) emerge when the pairing strength surpasses the gap, suggesting the possible presence of topologically nontrivial states.

To determine the exact topological properties of the Hamiltonian (\ref{eq2}), a topological invariant needs to be introduced. Due to the breaking of time-reversal symmetry by the magnetic field, the system possesses only particle-hole symmetry $C=\tau_{y}\sigma_{y}K$ ($K$ denotes the complex conjugate operator), rendering it a one-dimensional class D superconductor\cite{63}, which is characterized by a $\mathbf{Z}_{2}$ topological invariant associated with the particle-hole symmetry operator. The invariant, initially proposed by Kitaev\cite{62}, is given by the sign of the Pfaffian of the Hamiltonian matrix expressed in the Majorana fermion representation. In this model, the $\mathbf{Z}_{2}$ invariant is written as
\begin{eqnarray}\label{eq3}
M=Sgn\prod_{k=K_{0}}\{[C(k)+V]^{2}+J^{2}S^{2}-\Delta^{2}-\mu^{2}\},
\end{eqnarray}
with $C(k)=2t\sin\phi\cos k$ and $K_{0}=0,\pi$. A detailed derivation process is presented in Appendix A. The condition $M<0(M>0)$ indicates the existence of topologically nontrivial (trivial) states. We show the phase diagrams in Figs. \ref{fig2}(b)-(d). Therefore, we confirm that the zero modes observed in Fig. \ref{fig2}(a) are indeed Majorana zero modes, as they manifest within the topologically nontrivial regime denoted by the red dashed line in Fig. \ref{fig2}(c).

	\section{The influence of spin-orbit interaction}\label{Sec3}
The aforementioned model relies on the presence of AFM order, wherein the magnetic field must be oriented perpendicularly to the AFM moments.
As elucidated in Sec. \ref{Sec1}, there exist various means to attain AFM order. We concentrate on the RKKY mechanism and employ a classic spin approximation for the magnetic atoms, treating them as local magnetic fields, since many experimental results have closely matched theoretical calculations well under the approximation\cite{45,50,64,65,66,67,68,69}. Prior research has indicated that when a magnetic atomic chain is formed on an s-wave superconductor, the magnetic atoms self-organize into a helical magnetic moment structure, where the pitch angle is commensurate with the Fermi wave vector at 2$k_{F}a$\cite{21,22,23}. Nevertheless, the moments of an AFM order established via this method exhibit a random orientation. To fix the orientation, it is imperative to consider additional effects.

First, we consider the Rashba SOC induced in a magnetic atomic chain by breaking the inversion symmetry along the $z$ direction (assuming the substrate surface is in the $x$-$y$ plane) when the chain is deposited on a substrate. The Rashba SOC takes the form $\alpha_{R}(\mathbf{k}\times\mathbf{\sigma})_{z}$, which breaks the system's spin-rotation symmetry from SU(2) down to U(1), thereby providing a means to adjust the orientation of the AFM moments. Since our focus is not on discussing the topologically nontrivial superconductivity induced by a strong SOC in this model, we minimize the influence of the SOC on topological properties by assuming a negligible Rashba SOC with a tiny $\alpha_{R}$.
When the chain is oriented along the $y$ direction, the SOC becomes $(\alpha_{R}k,0,0)$, requiring the moments of the AFM order to be either parallel to the $x$ axis, lying in the $y$-$z$ plane, or canted along $x$ axis relative to the $y$-$z$ plane.

In our model, the AFM order is achieved by setting the chemical potential to $\mu=0$. Under periodic boundary conditions, the normal-state Hamiltonian with Rashba SOC can be expressed in $k$ space as
\begin{eqnarray}\label{eq4}
H^{\alpha}(k)&=&(2t\cos\frac{k}{2}+2\alpha_{R}\sin\frac{k}{2}\sigma_x)\Gamma_{k/2}+Js_{x}\sigma_x\Gamma_{i}\nonumber\\
&+&(Js_{y}\sigma_y+Js_{z}\sigma_z)\Gamma_z,
\end{eqnarray}
with $k\in(-\pi,\pi)$. $\Gamma_{i}$ represents either $\Gamma_{z}$ or the identity matrix, denoted as $\Gamma_{0}$. The possible orientations of the AFM moments are described by selecting different $\Gamma_{i}$. When $\Gamma_{i}=\Gamma_{z}$, the Hamiltonian describes a perfect AFM order with moments aligned along a random orientation, and the energy dispersions are given by
\begin{eqnarray}\label{eq5}
\epsilon_{12}^{\pm}(k)=\pm\sqrt{A^{2}+B^{2}\mp 2\sqrt{B^{2}(A^{2}-J^{2}s_{x}^{2})}},
\end{eqnarray}
with $A^{2}=4t^{2}\cos^{2}\frac{k}{2}+J^{2}S^{2}$ and $B=2\alpha_{R}\sin\frac{k}{2}$.
For the energy dispersions of the form $\epsilon_{1,2}^{\pm}(k)=\pm\sqrt{f\mp2\sqrt{g}}$, the two lower bands are given by $\epsilon_{1,2}^{-}(k)=-\sqrt{f\mp2\sqrt{g}}$. Since the system is half-filled, the two lower bands are completely occupied. At $T=0K$, the free energy of the system can be expressed as $G=\sum_{k}F(k)$ , where
$F(k)=\epsilon_{1}^{-}(k)+\epsilon_{2}^{-}(k)=-\sqrt{[\epsilon_{1}^{-}(k)+\epsilon_{2}^{-}(k)]^{2}}=-\sqrt{2}\sqrt{f+\sqrt{f^{2}-4g}}$.
Here, $F_{1}^{\alpha}(k)=-\sqrt{2A^{2}+2B^{2}+2\sqrt{(A^{2}-B^{2})^{2}+4B^{2}J^{2}s_{x}^{2}}}$. Evidently, with respect to $s_{x}$, the minimum value of $F_{1}^{\alpha}(k)$ occurs at $s_{x}=\pm S$ for every $k$.

When $\Gamma_{i}=\Gamma_{0}$, the Hamiltonian contains a canted AFM order.
The energy dispersions are
\begin{eqnarray}\label{eq6}
\epsilon_{1}^{\pm}(k)=\pm\sqrt{A^{2}-2aJs_{x}}-B,\\\nonumber
\epsilon_{2}^{\pm}(k)=\pm\sqrt{A^{2}+2aJs_{x}}+B,
\end{eqnarray}
with $a=2t\cos\frac{k}{2}>0$ for $k\in(-\pi,\pi)$. The sum of the two lower bands can take two forms. One is $F_{2}^{\alpha}(k)=-2|B|$, which is discarded since it is larger than $F_{1}^{\alpha}(k)$ for all values of $k$. The other form is $F_{2}^{\alpha}(k)=\epsilon_{1}^{-}(k)+\epsilon_{2}^{-}(k)=-\sqrt{2A^{2}+2\sqrt{A^{4}-4a^{2}J^{2}s_{x}^{2}}}$, whose minimum value occurs at $s_{x}=0$ for each $k$.

We find $F_{1min}^{\alpha}(k)=F_{1}^{\alpha}(k)|_{s_{x}=\pm S}<F_{2}^{\alpha}(k)|_{s_{x}=0}$=$F_{2min}^{\alpha}(k)$, indicating $F_{1min}^{\alpha}(k)<F_{2min}^{\alpha}(k)$ and the free energy $G_{1min}^{\alpha}<G_{2min}^{\alpha}$. Therefore, the system favors an AFM order with the maximum $|s_{x}|=S$ and minimum $|s_{y}|=|s_{z}|=0$, implying that the AFM moments align with the $x$-direction (noting that the chain is oriented along the $y$-direction). Obviously, if the chain is instead oriented along the $x$-direction with the SOC taking the form $(0,\alpha_{R}k,0)$, the AFM moments will preferentially align with the $y$-direction.
These results indicate that the Rashba SOC can fix the AFM moments to be confined within the surface plane and perpendicular to the ring, as illustrated in Fig. \ref{fig1}(a).

\section{The Effect of Magnetic Field}\label{Sec4}
\begin{figure}
\scalebox{1.0}{\includegraphics[width=0.45\textwidth]{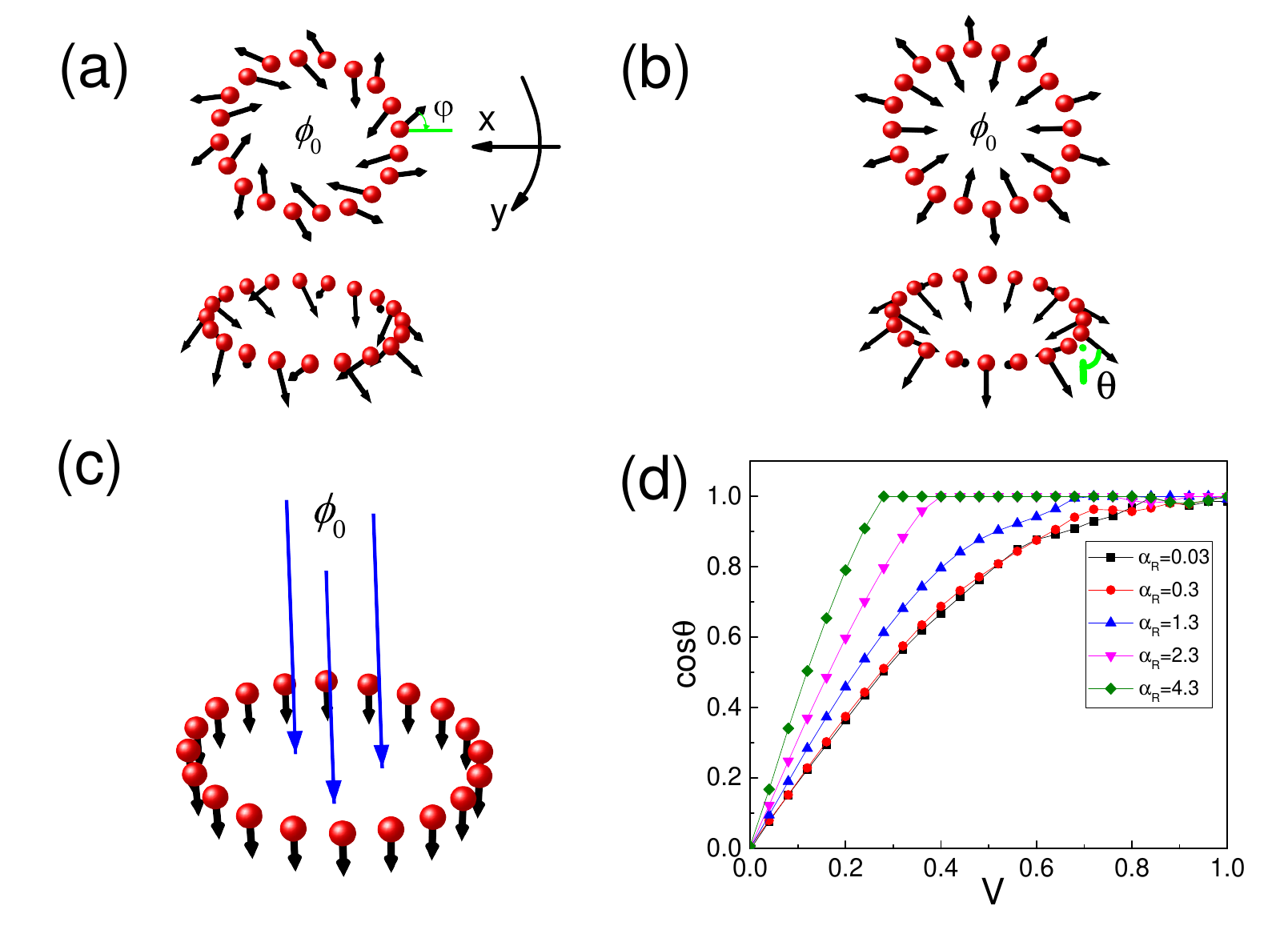}}
\caption{\label{fig3} (a) An AFM ring on an s-wave superconductor canted under a magnetic field $V$ applied perpendicular to the ring plane. $\varphi$ represents a random angle. (b) The canted AFM moments, with the AFM components along the $x$ direction and the FM components along the $z$ direction. (c) The FM moments along the $z$ direction, twisted from an initial AFM order by a large $V$. (d) The cosine of the canted angle as a function of $V$ for different SOC strengths with $JS=0.2$.}
\end{figure}

An external magnetic field also breaks the SU(2) spin-rotation symmetry down to U(1) spin-rotation symmetry. The perpendicular magnetic field applied on the surface requires the AFM moments to be either parallel to the $z$ axis, confined within the $x$-$y$ plane, or canted along the $z$ axis relative to the $x$-$y$ plane.
The normal-state Hamiltonian containing these three scenarios can be written as
\begin{eqnarray}\label{eq7}
H^{M}(k)&=&2t\cos\frac{k}{2}\Gamma_{k/2}+V\sigma_z\nonumber\\
&+&(Js_{x}\sigma_x+Js_{y}\sigma_y)\Gamma_z+Js_{z}\sigma_z\Gamma_i.
\end{eqnarray}

When $\Gamma_i=\Gamma_z$, the Hamiltonian describes the AFM orders encompassing the first and second scenarios. The sum of the two lower bands gives $F_{1}^{M}(k)=-\sqrt{2(A^{2}+V^{2})+2\sqrt{(A^{2}+V^{2})^{2}-4V^{2}(a^{2}+J^{2}s_{z}^{2})}}$. For each value of $k$, $F_{1}^{M}(k)$ is minimized when $s_{z}=0$, implying that the minimum value of the free energy [$G_{1}^{M}=\sum_{k}F_{1}^{M}(k)$] occurs at $s_{z}=0$.

When $\Gamma_{i}=\Gamma_{0}$, the system exhibits a FM order with the moments aligned with the direction of the magnetic Zeeman field $V$. The corresponding Hamiltonian describes the third scenario, with
$F_{2}^{M}(k)=-\sqrt{2f_{2}+2\sqrt{f_{2}^{2}-4g_{2}}}$, where $f_{2}=A^{2}+V^{2}+2VJs_{z}$ and $g_{2}=a^{2}(Js_{z}+V)^{2}$. When $V>a$, $F_{2}^{M}(k)$ monotonically decreases in the interval $[-S,S]$.
When $V<a$, $F_{2}^{M}(k)$ monotonically decreases in the interval $[-S,s_{z0}]$, with $s_{z0}=\frac{JS^{2}V}{a^{2}-V^{2}}>0$. The minimum value of $F_{2}^{M}(k)$ occurs at $s_z=s_{z0}$ when $s_{z0}<S$, and $s_z=S$ when $s_{z0}>S$. Thus, for both $V>a$ and $V<a$, a positive $s_{z}$ is required for the minimum value of $F_{2}^{M}(k)$. As a consequence, the minimum value of the free energy, $G_{2}^{M}=\sum_{k}F_{2}^{M}(k)$, must occur at a positive $s_{z}'$. Since the preceding discussion does not have any limitation on $s_{x}$ and $s_{y}$, if the condition $s_{z}'<S$ is satisfied, where $S^{2}=s_{x}^{2}+s_{y}^{2}+s_{z}'^{2}$, nonzero values for $s_{x}$ and $s_{y}$ are required. A detailed derivation is provided in Appendix B.

It is obvious that $G_{1min}^{M}=G_{1}^{M}|_{s_{z}=0}=G_{2}^{M}|_{s_{z}=0}>G_{2}^{M}|_{s_{z}'}=G_{2min}^{M}$, indicating that the system favors a canted AFM order, whose moments' out-of-plane components in the $z$ direction are ferromagnetically ordered and the in-plane components in the $x$-$y$ plane remain antiferromagnetically ordered, as shown in Fig. \ref{fig3}(a). Continuously increasing the strength of the magnetic field, the AFM order will eventually transition into a fully FM order, as illustrated in Fig. \ref{fig3}(c).

\section{Magnetism under spin-orbit coupling, magnetic field and superconductive pairing}\label{Sec5}
When both the Rashba SOC and perpendicular magnetic field are present simultaneously, they naturally compete due to their adherence to different U(1) spin-rotation symmetries. This competition yields a unilateral outcome, as the U(1) symmetry of the SOC is disrupted, while the U(1) symmetry of the magnetic field remains intact. Consequently, the magnetic field induces FM order, thereby breaking the U(1) symmetry of the SOC. The subsequent proof will substantiate this claim.

The Hamiltonian, which encompasses both a perpendicular magnetic field and in-plane Rashba SOC, is derived by incorporating $2\alpha_{R}\sin\frac{k}{2}\sigma_x\Gamma_{k/2}$ into Hamiltonian (\ref{eq7}).
When $\Gamma_{i}=\Gamma_{z}$, $F_{1}(k)$ yields $F_{1}(k)=-\sqrt{2f_{1}+2\sqrt{f_{1}^{2}-4g_{1}}}$, where $f_{1}=A^{2}+B^{2}+V^{2}$ and $g_{1}=B^{2}(A^{2}-J^{2}s_{x}^{2})+V^{2}(a^{2}+J^{2}s_{z}^{2})$. The minimum value of $F_{1}(k)$ coincides with the minimum of $g_{1}$, necessitating $s_{x}=\pm S$, and $s_{z}=s_{y}=0$.

When $\Gamma_{i}=\Gamma_{0}$, the equation
$F_{2}(k)=-\sqrt{2f_{2}+2\sqrt{f_{2}^{2}-4g_{2}}}$ holds, with $f_{2}=A^{2}+V^{2}+B^{2}+2VJs_{z}$ and $g_{2}=B^{2}(a^{2}+J^{2}s_{y}^{2})+a^{2}(Js_{z}+V)^2$.
It is evident that to obtain the minimum value of $F_{2}(k)$, $s_{y}=0$ is required. When $V>aT$, where $T=\frac{Js_{z}+V}{\sqrt{B^2+(Js_{z}+V)^2}}$, $F_{2}(k)$ monotonically decreases within the interval $[-S,S]$. When $V<aT$, $F_{2}(k)$ exhibits monotonic decrease in the interval $[-S,s_{z0}]$, with $s_{z0}=V\frac{D-2(a^2-V^2)+\sqrt{D^2+4B^2(a^2-V^2)}}{2J(a^2-V^2)}>V\frac{B^2+J^2S^2}{2J(a^2-V^2)}>0$ and $D=a^2-V^2+B^2+J^2S^2$. The minimum value of $F_{2}(k)$ occurs at $s_z=s_{z0}$ when $s_{z0}<S$, and at $s_z=S$ when $s_{z0}>S$. These results closely resemble those obtained when considering only the perpendicular magnetic field, as discussed in Sec. \ref{Sec4}. The minimum value of $F_{2}(k)$ requires a positive $s_{z}$ under both conditions $V>aT$ and $V<aT$. Hence, $G_{2min}$ is located at a positive $s'_{z}$. For a more comprehensive derivation, refer to Appendix C.

It is observed that $G_{1min}=G_{1}|_{s_z=0}=G_{2}|_{s_z=0}>G_{2}|_{s_z=s'_z}=G_{2min}$, indicting a preference for canted antiferromagnetically ordered moments with induced FM components aligned parallel to the magnetic field $V$. Furthermore, the SOC continues to influence the orientation of the AFM components, as depicted in Fig. \ref{fig3}(b), rather than allowing random orientations, which is indicated by the requirement of $s_{y}=0$ for $G_{2min}$.
Figure \ref{fig3}(d) presents the numerically solved results of the induced FM orders under varying strengths of SOC and magnetic field within a 160-site ring. It appears that the SOC not only influences the orientation of the AFM components, but also plays a role of compensating for the effects of the magnetic fields.

When introducing a flux $\phi_{0}$ into the system, the wave vector $k$ is replaced with $k\pm2\phi$ (where $\phi=\phi_{0}/L$), while the structure of $F(k)$ remains unchanged. The flux solely shifts the bands along the $k$ direction without elevating or lowering any band and has no impact on the free energy.
\begin{figure}
\scalebox{1.0}{\includegraphics[width=0.45\textwidth]{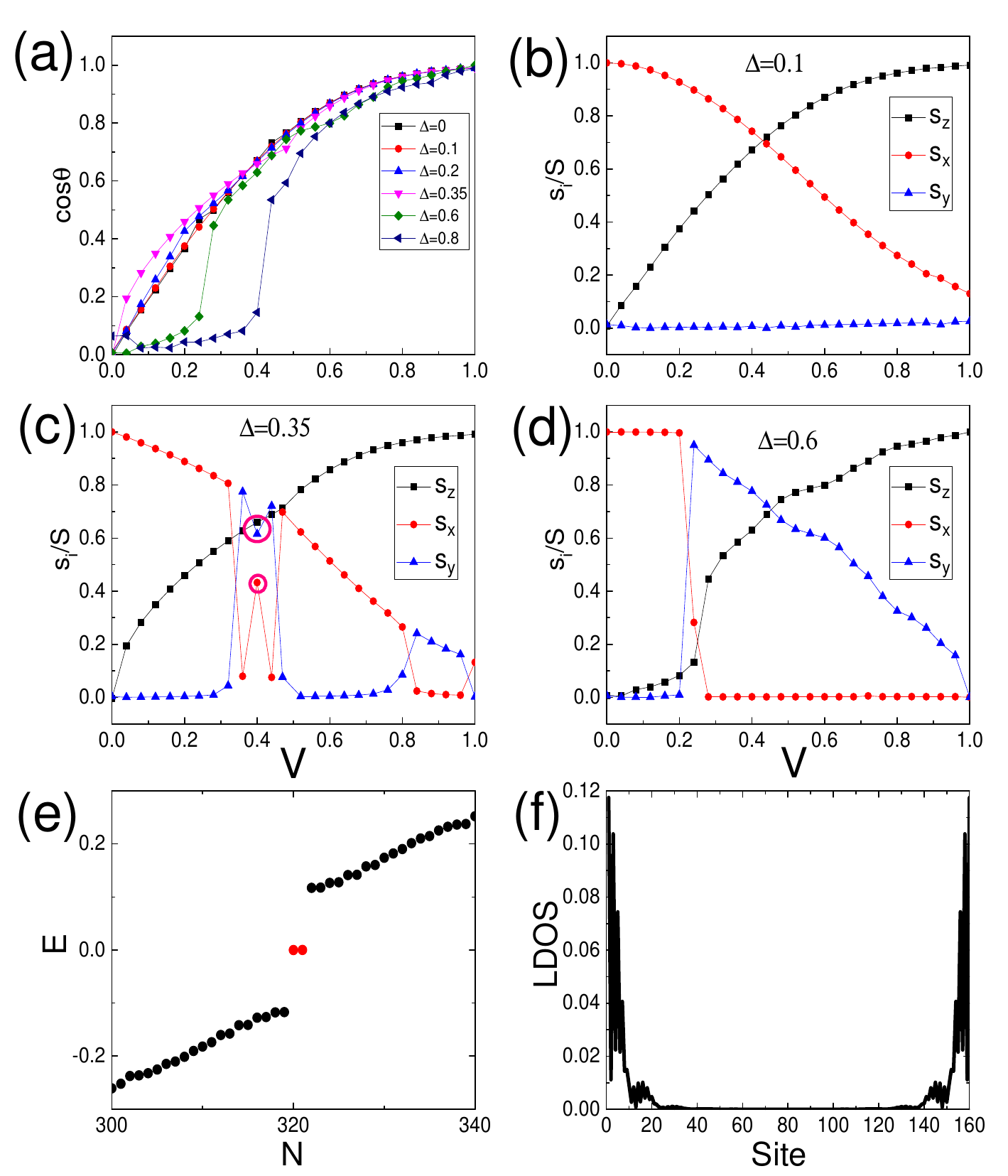}}
\caption{\label{fig4}Self-consistently solved magnetic orders and the corresponding zero modes in real space for a system with 160 sites. (a) The FM moment components induced by a magnetic field; (b)-(d) both the induced FM and the in-plane AFM components along the $x$-$y$ directions under different superconducting pairing strengths. The parameters are set as $\alpha_{R}=0.05$, $JS=0.2$, $\phi=0.05\pi$ and $\mu=0$. (e),(f) The open boundary energy levels of the ring and LDOS of one zero mode, respectively. The magnetic configuration is selected from (c) and marked by open circles for $\Delta=0.35$ and $V=0.4$. The ``N" in (e) denotes the energy-level index.}
\end{figure}

When the AFM ring is deposited on an s-wave superconductor, the canted magnetic moments persist as AFM in the $x$-$y$ plane and FM along the $z$ direction.
We self-consistently solve the real-space Bogoliubov-de Gennes (BdG) equation for the modified Hamiltonian (\ref{eq1}), where $J(-1)^{j}\mathbf{S}_{j}$ is substituted with $J\mathbf{S}_{j}$ and $\mathbf{S}_{j}=S(\sin\theta_{j}\cos\varphi_{j},\sin\theta_{j}\sin\varphi_{j},\cos\theta_{j})$. We select a ring consisting of 160 sites and initialize the process with random values of $\theta_{j}$ and $\varphi_{j}$ at each site. Throughout this process, we exclusively perform self-consistent calculations to determine the orientation of each magnetic moment with a fixed pairing strength $\Delta$ (the detailed self-consistency process can be found in Ref. \cite{22}). Figure \ref{fig4}(a) presents numerical results of the induced FM components of the magnetic moments for varying pairing strengths, while more detailed results, including AFM components, are depicted in Figs. \ref{fig4}(b) and 4(c).
Our results indicate that a weak pairing, such as $\Delta=0.1$, does not disrupt the magnetic orders present in the normal state ($\Delta=0$), as illustrated in Fig. \ref{fig4}(a). When the pairing strength increases, orders deviate from the normal state. This deviation generates not only in the induced FM components, but also in the AFM components. For example, when $\Delta=0.35$, the AFM components no longer align along the $x$ axis but can potentially exist along any direction within the $x$-$y$ plane. With a larger $\Delta$, such as $\Delta=0.6$, the AFM components are completely oriented along the $y$ axis. These deviations occur due to the significant Cooper pairing strongly distorting the band structure, leading to adjustments in the free energy and magnetic order. Nevertheless, as long as the AFM components are confined to the $x$-$y$ plane, The Hamiltonian (\ref{eq1}) can be realized by modifying $Js_{x(y)}$ to $Js_{x(y)}\sin\theta$ and $V$ to $V+JS\cos\theta$. Figures. \ref{fig4}(e) and 4(f) display the Majorana zero modes and one example of their local density of states (LDOS) in a scenario where the AFM components deviate from the $x$ direction. The magnetic structure, selected from Fig. \ref{fig4}(c), is marked by open circles with fixed $\Delta=0.35$ and $V=0.4$.

\section{Magnetic Order on 2D Surface}\label{Sec6}
\begin{figure}
\scalebox{1.0}{\includegraphics[width=0.45\textwidth]{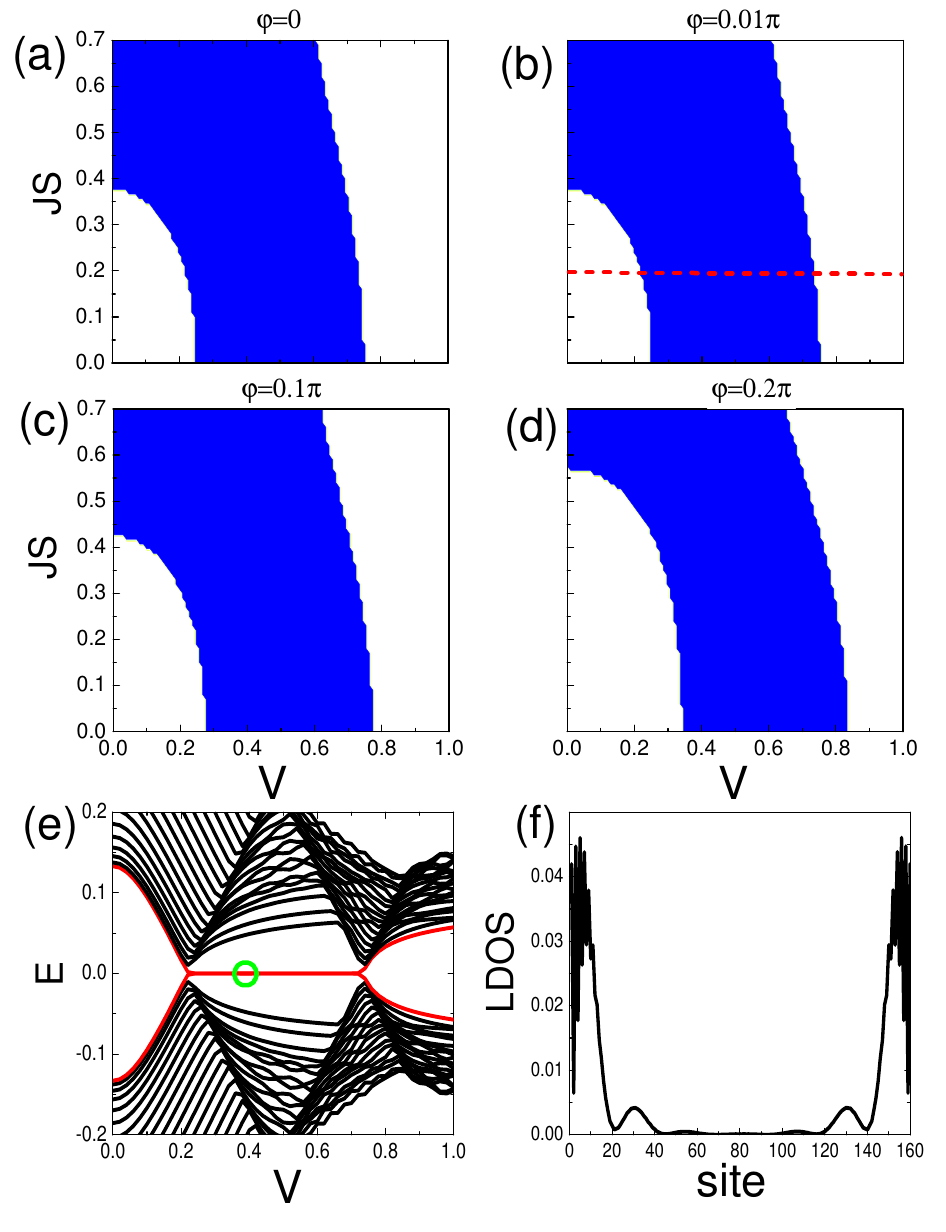}}
\caption{\label{fig5} (a)-(d) The topologically nontrivial phases (blue regions) in the parameter space with different patch angles $\varphi$. (e) The open boundary energy spectrum with the scanning trajectory denoted by a red dashed line in (b). (f) The LDOS of the zero mode marked in (e) by the open circle. In (a)-(f) , $\alpha_{R}=0$, $\phi=0.04\pi$, and $\Delta=0.5$. Other parameters in (a)-(d) are $\varphi=0$, $0.01\pi$, $0.1\pi$ and $0.2\pi$, respectively. In (e) and (f), $\varphi=0.01\pi$, $JS=0.2$, and in (f), $V=0.4$.}
\end{figure}

If the substrate superconductor surface is sufficiently large to be treated as two dimensional (2D), the magnetic atoms exhibit a preference for FM order \cite{41}. We find that topological superconductivity is reconstructible by adopting the magnetic atoms on the next-nearest-neighbor lattice sites. When a vertical magnetic field is applied to the surface and the substrate's influence is integrated out\cite{21}, the 1D effective Hamiltonian can be expressed in $k$ space as
\begin{eqnarray}\label{eq8}
\widehat{H}(k)&&=[\xi_{0}(k)\tau_z+\eta_{0}(k)]\Gamma_{k/2}+\Delta\tau_{x}+V\sigma_z\nonumber\\
&&+J(s_{x}\sigma_{x}+s_{y}\sigma_y)(\Gamma_{z}+\Gamma_{0})/2.
\end{eqnarray}
The system remains a one-dimensional class-D superconductor, and the $\mathbf{Z}_{2}$ invariant can be determined by calculating the Pfaffian of the Hamiltonian matrix in the Majorana fermion representation. The calculation method is detailed in Appendix A and the corresponding topological phase diagram is presented in Fig. \ref{fig5}(a).

Contrary to AFM orders, the SOC or magnetic field do not ensure the perpendicularity between the ferromagnetically ordered moments and the magnetic field, which is crucial for the system to exhibit topological nontriviality. The U(1) spin-rotation symmetry of the Rashba SOC promotes FM order with in-plane moments, but it is disrupted by the perpendicular magnetic field, favoring FM order with moments parallel to the field. When both the magnetic field and SOC are present simultaneously, the FM moments are likely to align parallel to the magnetic field.
However, previous studies have shown that in the presence of SOC, magnetic atoms favor a helical magnetic order with a pitch angle of $2k_{R}a$, where $k_{R}=m^{*}\alpha_{R}$\cite{41,23}, and $m^{*}$ represents the effective mass of electrons.
When the magnetic field exceeds the critical value associated with the SOC energy scale $\alpha_{R}k_{F}$, a homogeneous magnetization parallel to $\mathbf{B}$ is induced and the helical components become perpendicular to $\mathbf{B}$\cite{23}. In the subsequent discussion, we aim to briefly explore the potential for achieving topological superconducting states in this helically ordered ring. For simplicity, we exclude the SOC term in the Hamiltonian and disregard the induced homogeneous magnetization along $\mathbf{B}$. This decision is justified by the minor influence of the tiny SOC on the topological properties, resulting solely in a small pitch angle for the helical order. Moreover, the induced homogeneous magnetization is relatively small in comparison to the external field, and its magnitude varies with different strengths of $\mathbf{B}$. After considering the negligible SOC and the resulting small pitch angle, denoted as $\varphi$, we apply the transformation $d_{j\uparrow}=c_{j\uparrow}e^{ij\varphi/4}$ and $d_{j\downarrow}=c_{j\downarrow}e^{-ij\varphi/4}$ to the Hamiltonian, yielding the expression of the Hamiltonian in $k$ space as
\begin{eqnarray}\label{eq9}
\widehat{H}'(k)&&=[\xi_{k/2}\tau_z+\eta_{k/2}]\Gamma_{k/2}+\Delta\tau_{x}+V\sigma_z\nonumber\\
&&+JS\sigma_{x}(\Gamma_{z}+\Gamma_{0})/2,
\end{eqnarray}
with $\xi_{k/2}=2t\cos\phi\cos\frac{\varphi}{4}\cos\frac{k}{2}+2t\sin\phi\sin\frac{\varphi}{4}\cos\frac{k}{2}\tau_z\sigma_z$ and $\eta_{k/2}=2t\sin\phi\cos\frac{\varphi}{4}\sin\frac{k}{2}-2t\cos\phi\sin\frac{\varphi}{4}\sin\frac{k}{2}\tau_z\sigma_z$.
It can be anticipated that if the pitch angle is small, the system's topological properties closely resemble those of a ring with in-plane FM moments ($\varphi=0$).
In Fig. \ref{fig5}(b), we present the topological phase diagram of the Hamiltonian (\ref{eq9}) with a small pitch angle of $\varphi=0.01\pi$. The topologically nontrivial area is almost the same as that of the FM order. To test the stability of the topological phases against the helical orders, we examine the effects of different pitch angles and present the results in Figs. \ref{fig5}(b)-5(d). The comparisons confirm that the topological phases are robust despite variations in the pitch angles. Since a half-filled ring with a helical order generally does not exhibit topologically nontrivial superconductivity, these nontrivial phases are induced by the flux $\phi_{0}$. Finally, we show the open boundary spectrum of the ring with 160 sites in Fig. \ref{fig5}(e), along with the LDOS of a zero mode in Fig. \ref{fig5}(f). This zero mode is localized at the ends of the ring and identified as a Majorana zero mode.

	\section{Summary and Discussion}\label{Sec7}
	We present a one-dimensional topological superconductor model with an AFM ordered ring on an s-wave superconductor. The model requires a perpendicular magnetic field imposing on the ring, providing both a Zeeman field normal to the AFM moments and a magnetic flux through the ring. We first analyze the effects of an in-plane Rashba SOC and the perpendicular magnetic field in the normal states. Both of them reduce the symmetry of the system, leading to distinct orientations of the magnetic moments. The Rashba SOC favors in-plane and chain-perpendicular AFM moments, while the perpendicular magnetic field cants the in-plane AFM moments by inducing FM components parallel to the field and maintaining the in-plane AFM components with a random direction.
When both the in-plane Rashba SOC and perpendicular magnetic field are present, the competition between their influences is one sided.
The magnetic field tends to induce the FM components along its direction, while the SOC reinforces the effect of the magnetic field and governs the orientation of the in-plane AFM components by ensuring that they are perpendicular to the chain.

Second, we investigate the influence of the superconducting pairing which distorts the bands near the Fermi level. The pairing primarily impacts the canted AFM order by twisting the orientation of the in-plane moment components, which appears to change from being perpendicular to the ring to being parallel as the pairing strength increases. Such a twist does not hinder the realization of one-dimensional topological superconductivity.

Third, on the two-dimensional surface, the RKKY effect favors FM magnetic order, whose moments cannot be oriented normal to the external magnetic field. We introduce a minor Rashba SOC to construct a helical mgnetic order with a small pitch angle. The helical magnetic order's moments can be oriented in-plane by applying a perpendicular magnetic field. With this magnetic order, if magnetic atoms are deposited on next-nearest-neighbor lattice sites of an s-wave superconductor, a magnetic flux can induce topological superconducting states.

To achieve topological nontriviality in our model, $J^2S^2-\mu^2$ and $\Delta^2$ must be on the same order of magnitude, as illustrated in Fig. \ref{fig2}(b). In experiments, the exchange coupling strengths are expected to be of the order of electronvolts, while the superconducting order parameters $\Delta$ are of the order of millielectronvolts when depositing transition-metal atoms onto conventional superconductors. It is a big challenge to find high $\Delta$ superconductors, while increasing the chemical potentials and reducing the strengths of exchange couplings are feasible solutions. There are reports of achieving small $J$ when depositing Fe atoms onto the surface of superconducting $Ta(100)-(3\times3)O$\cite{70}, and also reports of reducing $J$ through the hydrogenation of adatoms\cite{71} or replacing the magnetic atoms with paramagnetic metal-organic molecules, where the molecular ligand with inert organic groups separates the central magnetic ion from its conducting environment\cite{72}.

Recent studies have investigated the impact of quantum many-body effects on the topological properties of dilutely deposited magnetic chains, where the spins of the magnetic atoms are treated quantum mechanically\cite{73,74}. It is a challenge to stabilize the magnetic structure under this effect. However, the actual spin structure of a magnetic chain is influenced by various effects, such as the RKKY effect, Dzyaloshinskii-Moriya interaction, SOC, and the Kondo effect. The dominant interaction depends on the substrate material, adatom species, distances between magnetic atoms, and exchange coupling strengths. Nevertheless, the classical approximation remains a valuable tool, given its success in explaining numerous experimental observations\cite{45,50,64,65,66,67,68,69}. Our proposal presents a potential approach to engineer the magnetic spin structure and achieve topological superconductivity. We believe this manuscript may provide valuable insights for experimental investigations in this field.

    \section{Acknowledgments}\label{Sec8}
	This work was supported by the National Natural Science Foundation of China (Grant Nos. 11947082, 12375022, 12065011, 11975110), the Scientific and Technological Research Fund of Jiangxi Provincial Education Department (Grant No. GJJ190577), and Zhejiang Sci-Tech University Scientific Research Start-up Fund (Grant No. 20062318-Y).
\onecolumngrid
	\section*{Appendix A: Calculation of the invariant $M$}\label{Sec9}
Here we present two approaches to calculate the $\mathbf{Z}_{2}$ invariant $M$ of the Hamiltonian (\ref{eq1}).
The first approach involves transforming the Hamiltonian directly into the Majorana fermion representation. We choose the basis $[r_{A,j},r_{B,j}]^T$, where $r_{\delta,j}= [a_{\delta,2j-1,\uparrow},a_{\delta,2j,\uparrow},a_{\delta, 2j-1,\downarrow},a_{\delta,2j,\downarrow}]$ and $c_{\delta,j,\alpha}=\frac{1}{\sqrt{2}}(a_{\delta, 2j-1,\alpha}+i a_{\delta, 2j,\alpha})$, $c^{\dag}_{\delta,j,\alpha}=\frac{1}{\sqrt{2}}(a_{\delta, 2j-1,\alpha}-i a_{\delta, 2j,\alpha})$, with $\delta=A/B$ representing the sublattice sites. Subsequently, we perform a Fourier transformation to express the Hamiltonian in the Majorana fermion representation in $k$ space,
\begin{eqnarray}\label{eqA1}
\widehat{H}_{0}(k)=(2t\sin\phi\sin\frac{k}{2}-2t\cos\phi\cos\frac{k}{2}\sigma_y)\Gamma_{k/2}-\mu\sigma_y-V\tau_z\sigma_y+\Delta\tau_y\sigma_x-(Js_{x}\tau_x\sigma_y-Js_{y}\tau_y)\Gamma_z.\nonumber
\end{eqnarray}
The $\mathbf{Z}_{2}$ invariant is expressed as
\begin{eqnarray}\label{eqA2}
M=Sgn\prod_{k=K_{0}}\{Pf[\widehat{H}_{0}(k)]\}=Sgn P(0)P(\pi),\nonumber
\end{eqnarray}
with $Pf$ denoting the Pfaffian and $K_{0}=0$ or $\pi$, $P(0)=(4t^2\cos^2\phi+\Delta^2+J^2S^2-\mu^2-V^2)^2-4(J^2S^2-\mu^2)(\Delta^2-V^2)$ and $P(\pi)=(4t^2\sin^2\phi-\Delta^2+J^2S^2-\mu^2-V^2)^2-4V^{2}(\Delta^{2}-J^{2}S^{2}+\mu^{2})$.

Employing the second approach, we initially perform a unitary transformation on the Hamiltonian (\ref{eq1}), such that $c_{j\uparrow}=e^{i\frac{\varphi_j}{2}}d_{j\uparrow}$ and $c_{j\downarrow}=e^{-i\frac{\varphi_j}{2}}d_{j\downarrow}$. Here, $\varphi_{j+1}-\varphi_{j}=\pi$ represents the different angle between the magnetic moments of two neighboring atoms in the antiferromagnetically ordered magnetic system. After that, the Hamiltonian can be written as
\begin{eqnarray}\label{eqA3}
H_{0}'=\sum_{j}ite^{-i\phi}d_{j}^{\dag}\sigma_zd_{j+1}+Jd_{j}^{\dag}\mathbf{S}\cdot\boldsymbol{\sigma}d_{j}
+\mu d_{j}^{\dag}d_{j}+V d_{j}^{\dag}\sigma_zd_{j}+\Delta d_{j\uparrow}^{\dag}d_{j\downarrow}^{\dag}+h.c.,\nonumber
\end{eqnarray}
with $d_{j}^{\dag}=(d_{j\uparrow}^{\dag},d_{j\downarrow}^{\dag})$. Then we express the transformed Hamiltonian in the Majorana fermion representation. We employ the basis $[a_{2j-1,\uparrow},a_{2j,\uparrow},a_{ 2j-1,\downarrow},a_{2j,\downarrow}]^{T}$ to the transformations $d_{j,\alpha}=\frac{1}{\sqrt{2}}(a_{ 2j-1,\alpha}+i a_{2j,\alpha})$ and $d^{\dag}_{j,\alpha}=\frac{1}{\sqrt{2}}(a_{ 2j-1,\alpha}-i a_{2j,\alpha})$ to express $H_{0}'$ in the Majorana fermion representation and then express it in $k$ space. We have
\begin{eqnarray}\label{eqA4}
\widehat{H}_{0}'(k)=2t\cos\phi\sin k\tau_z-(2t\sin\phi\cos k+V)\tau_z\sigma_y-\mu\sigma_y+\Delta\tau_y\sigma_x-Js_{x}\tau_x\sigma_y+Js_{y}\tau_y.\nonumber
\end{eqnarray}
The $\mathbf{Z}_{2}$ invariant is written as
\begin{eqnarray}\label{eqA5}
M=Sgn\prod_{k=K_{0}}\{Pf[\widehat{H}_{0}'(k)]\}=Sgn\prod_{k=K_{0}}\{[C(k)+V]^{2}+J^{2}S^{2}-\Delta^{2}-\mu^{2}\},\nonumber
\end{eqnarray}
with $C(k)=2t\sin\phi\cos k$ and $K_{0}=0,\pi$.

We have confirmed the consistency of the results obtained by the two methods by comparing the phase diagrams calculated using each method.

	\section*{Appendix B: Derivation of the influence of magnetic field}\label{Sec10}
When $\Gamma_{i}=\Gamma_{z}$, the energy dispersions of the Hamiltonian (\ref{eq7}) are given by
\begin{eqnarray}\label{eqA6}
\varepsilon_{1,2}^{\pm}(k)=\pm\sqrt{A^{2}+V^{2}\mp2\sqrt{(a^{2}+J^{2}s_{z}^{2})V^{2}}}.\nonumber
\end{eqnarray}
We have $F_{1}^{M}(k)=-\sqrt{2(A^{2}+V^{2})+2\sqrt{(A^{2}+V^{2})^{2}-4V^{2}(a^{2}+J^{2}s_{z}^{2})}}$.
$F_{1}^{M}(k)$ attains its minimum value for all values of $k$ when $s_z=0$. Thus the free energy $G_{1}^{M}=\sum_{k}F_{1}^{M}(k)$ is minimized when $s_z=0$.

When $\Gamma_{i}=\Gamma_{0}$, the energy dispersions become
\begin{eqnarray}\label{eqA7}
\varepsilon_{1,2}^{\pm}=\pm\sqrt{A^{2}+V^{2}+2Js_{z}V\mp2a(Js_{z}+V)}\nonumber
\end{eqnarray}
and $F_{2}^{M}(k)=\varepsilon_{1}^{-}+\varepsilon_{2}^{-}$.
The extremum of $F_{2}^{M}(k)$ can be determined by differentiating $F_{2}^{M}(k)$ with respect to $s_z$.  $\partial F_{2}^{M}/\partial s_{z}=J\frac{(V-a)\varepsilon_{2}^{-}+(V+a)\varepsilon_{1}^{-}}{\varepsilon_{1}^{-}\varepsilon_{2}^{-}}$. When $a<V$, $\partial F_{2}^{M}/\partial s_{z}<0$, $F_{2}^{M}$ monotonically decreases. The minimum value of $F_{2}^{M}$ occurs at $s_{z}=S$. When $a>V$, $F_{2}^{M}$ has only one extremum value, which exists at $s_{z0}=\frac{JS^{2}V}{a^{2}-V^{2}}>0$, indicating that $F_{2}^{M}$ is monotonic in both sides of the extremum value. We can judge whether it monotonically decreases or increases through checking the sign of differentiating $F_{2}^{M}(k)$ at any arbitrary $s_z$. For convenience, we choose $s_{z}=-S$. We find when $JS<a+V$, $\partial F_{2}^{M}/\partial s_{z}|_{-S}=-\frac{2aJ^2S}{\varepsilon_{1}^{-}\varepsilon_{2}^{-}}<0$, and when $JS>a+V$, $\partial F_{2}^{M}/\partial Js_{z}|_{-S}=-2J\frac{JSV+a^{2}-V^{2}}{\varepsilon_{1}^{-}\varepsilon_{2}^{-}}<0$. $F_{2}^{M}$ monotonically decreases from $-S$ to $s_{z0}$. If $s_{z0}<S$, the minimum value of $F_{2}^{M}$ occurs at $s_{z}=S_{z0}$. If $s_{z0}>S$, the minimum value occurs at $S$. Since $s_{z0}$ is dependent on $k$, the precise value of the induced FM order cannot be readily determined. However, it can be affirmed with certainty that the free energy $G_{2}^{M}=\sum_{k}F_{2}^{M}(k)$ possesses a minimum value at a positive value of $s_{z}'$.

	\section*{Appendix C: Derivation of the influence when magnetic field and SOC coexist}\label{Sec11}
In Sec. \ref{Sec5}, when $\Gamma_{i}=\Gamma_{z}$, the dispersions become
\begin{eqnarray}\label{eqA8}
\varepsilon_{1,2}^{\pm}=\pm\sqrt{A^{2}+B^{2}+V^{2}\mp2\sqrt{B^{2}(A^{2}-J^{2}s_{x}^{2})+V^{2}(a^{2}+J^{2}s_{z}^{2})}}.\nonumber
\end{eqnarray}
$F_{1}(k)=-\sqrt{2f_{1}+2\sqrt{f_{1}^{2}-4g_{1}}}$ with $f_{1}=A^{2}+B^{2}+V^{2}$ and $g_{1}=B^{2}(A^{2}-Js_{x}^{2})+V^{2}(a^{2}+J^{2}s_{z}^{2})$. When $|s_{x}|=S$, $s_{z}=s_{y}=0$, $F_{1}(k)$ attains its minimum value for all values of $k$. Consequently, the free energy $G_{1}=\sum_{k}F_{1}(k)$ also possesses a minimum value when the condition $s_{x}=S$ is satisfied.

When $\Gamma_{i}=\Gamma_{0}$, the energy dispersions are given by
\begin{eqnarray}\label{eqA9}
\varepsilon_{1,2}^{\pm}=\pm\sqrt{f_{2}\mp2\sqrt{g_{2}}}=\pm\sqrt{(A^{2}+B^{2}+V^{2}+2VJs_{z})\mp2\sqrt{B^{2}(a^{2}+J^{2}s_{y}^{2})+a^{2}J^{2}s_{z}^{2}+2a^{2}VJs_{z}+a^{2}V^{2}}}. \nonumber
\end{eqnarray}
$F_{2}(k)=-\sqrt{2f_{2}+2\sqrt{f_{2}^{2}-4g_{2}}}$.
To obtain the minimal value of $F_{2}(k)$, the condition $s_{y}=0$ must be satisfied. Subsequently, the minimum value of $F_{2}(k)$ can be determined by differentiating it with respect to $s_z$. $\partial F_{2}/\partial s_{z}=J\frac{(V-aT)\varepsilon_{2}^{-}+(V+aT)\varepsilon_{1}^{-}}{\varepsilon_{1}^{-}\varepsilon_{2}^{-}}$ with $T=\frac{Js_{z}+V}{\sqrt{B^2+(Js_{z}+V)^2}}$ and $|T|<1$.

When $V<aT$, we have $Js_z+V>0$ and $V<a$. Let $\partial F_{2}/\partial s_{z}=0$; we get the extremum point $s_{z0}=V\frac{D-2(a^2-V^2)+\sqrt{D^2+4B^2(a^2-V^2)}}{2J(a^2-V^2)}>0$ with $D=a^2-V^2+B^2+J^{2}S^2$. We figure out the monotonicity in either side of the extremum point $s_{z0}$ via comparing the values of $(V-aT)\varepsilon_{2}^{-}$ and $(V+aT)\varepsilon_{1}^{-}$.
$[(V-aT)\varepsilon_{2}^{-}]^2-[(V+aT)\varepsilon_{1}^{-}]^2=\frac{1}{g_{2}}[V\sqrt{g_{2}}-a^2(Js_{z}+V)]^2(f_{2}+2\sqrt{g_{2}})
-\frac{1}{g_{2}}[V\sqrt{g_{2}}+a^2(Js_{z}+V)]^2(f_{2}-2\sqrt{g_{2}})=\frac{4}{\sqrt{g_{2}}}(C1-C2)$, with $C1=V^2g_{2}+a^4(Js_{z}+V)^2$ and $C2=a^2V(Js_{z}+V)f_{2}$. $C1-C2|_{s_{z}=-S}=a^2JS[a^2(JS-V)+V(JS-V)^2+VB^2]=\frac{a^2JS}{T^2}[V(V-JS)-(aT)^2]<\frac{a^2JS}{T^2}[V^2-(aT)^2]$. Here $T=T|_{s_z=-S}$.  We remind readers that here we are discussing the situation with $0<V<aT$. Therefore, $C1-C2<0$ and $|(V-aT)\varepsilon_{2}^{-}|<|(V+aT)\varepsilon_{1}^{-}|$. Finally, $\partial F_{2}/\partial s_{z}|_{s_{z}=-S}=\frac{(V-aT)\varepsilon_{2}^{-}+(V+aT)\varepsilon_{1}^{-}}{\varepsilon_{1}^{-}\varepsilon_{2}^{-}}<0$. Therefore, $\partial F_{2}/\partial s_{z}<0$ in the interval $[-S,s_{z0})$. $F_{2}(k)$ decreases from $s_{z}=-S$ to $s_{z0}$. The minimum value of $F_{2}(k)$ occurs at $s_{z0}$.

When $V>aT$, $\partial F_{2}/\partial s_{z}<0$, indicating that $F_{2}$ monotonically decreases in the interval $[-S,S]$. $F_{2}$ attains its minimum value when $s_z=S$.

Therefore, either $V<aT$ or $V>aT$, and the minimum value of $F_{2}(k)$ occurs at a positive $s_{z}$.

%\bibliography{ref1}
\twocolumngrid

\end{document}